# Bending short DNAs as transversely isotropic rings in series

Chenyu Shi[1], Meicheng Yao[1], Bin Chen[1,2*]

[1]Department of Engineering Mechanics, Zhejiang University, Hangzhou, People's Republic of China

[2]Key Laboratory of Soft Machines and Smart Devices of Zhejiang Province, Hangzhou, People's Republic of China

*To whom correspondence should be addressed: chenb6@zju.edu.cn

**Abstract**

Despite the significance of the high flexibility exhibited by short DNAs, there remains an incomplete understanding of their anomalous persistence length. In this study, we propose a novel approach wherein each fundamental characteristic of gene sequences within short DNAs is modeled as a transversely isotropic ring. Our comprehensive model analysis not only successfully replicates the observed high flexibility of short DNAs but also sheds light on the impact of sequence dependence, aligning with experimental findings. Furthermore, our analysis suggests that the bending behavior of short DNAs can be effectively described by the Timoshenko beam theory, accounting for shear considerations.

Flexibility of DNAs plays a crucial role in various cellular functions. A more flexible DNA is easier to cyclize, facilitating gene storage, gene replication, gene transcription, etc. [1,2]. High flexibility of DNAs influences gene expression by efficiently wrapping around the histone octamer [3]. A flexible DNA can form a ring structure to impede further transcription [4]. Moreover, the flexibility of short DNAs has fueled advancements in various biotechnological applications [5], with significant progress made in DNA origami for constructing materials with novel properties [6].

Traditionally, DNA has been regarded as a homogeneous and isotropic elastic rod, with its force-stretch relationship described by the worm-like chain (WLC) model [7,8]. However, recent advancements in molecular techniques have revealed that short DNAs exhibit much higher flexibility compared to long DNAs [9-11], rendering the classical WLC model inadequate in explaining the observed flexibility of short DNAs. In an effort to address this discrepancy, Wiggins et al. [12] demonstrated that kinks could significantly enhance the cyclization rate of short DNAs [13], while Yan et al. [14] showed that single-stranded temporary bubbles could also contribute to their flexibility. Although there have been debates [16] over whether the increased flexibility of short DNAs is attributed to kinks, bubbles, or anharmonic elasticity [15], Chen and Dong [17] introduced an anisotropic texture into the structural model of short DNAs, providing an explanation for their high flexibility without considering defects.

The sequence of base pairs has also been found to influence DNA flexibility [18-21]. Shon et al. [20] demonstrated that higher CG/GC content resulted in increased bending stiffness of short DNAs. Geggier et al. [21], through the design of DNA sequences with varying persistence lengths, confirmed the sequence dependence of DNA flexibility. Utilizing FRET (Fluorescence Resonance Energy Transfer) techniques, it was clearly demonstrated that the sequence of short DNAs directly impacts their flexibility [19]. Various theories have been proposed to explain the sequence dependence of short DNA flexibility, focusing on the differentiation of different base pairs at the individual level [16, 22]. However, it should be noted that most existing models are currently unable to simultaneously explain both the high flexibility of short DNAs and the influence of base pair sequence on their flexibility.

To address this issue, building upon the concept of anisotropic texture introduced by Chen and Dong [17], we propose a novel approach whereby each fundamental gene invariant within short DNAs is modeled as a transversely isotropic ring. Notably, our model predictions exhibit an exceptional level of agreement with experimental observations. Furthermore, our comprehensive analysis demonstrates that the anomalous flexibility of short DNAs can be effectively described by the Timoshenko beam theory, accounting for shear considerations. This study emphasizes the critical role of structural heterogeneity and anisotropy in determining the flexibility of short DNAs.

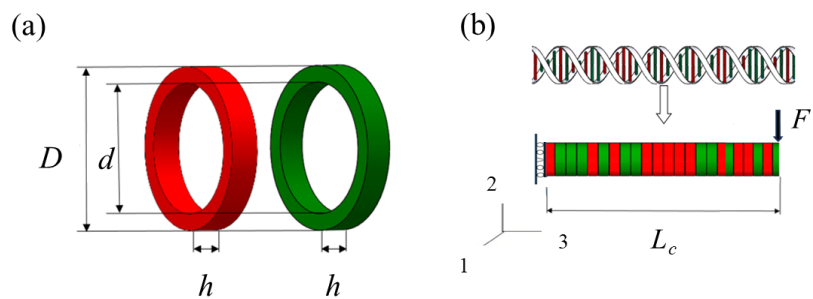

**Fig. 1** (a) Red and green rings represent the sequences of CG/GC and AT/TA, respectively. $D$, $d$, and $h$ denote the outer diameter, inner diameter, and height of a ring, respectively; (b) A short DNA with a contour length of $L_c$ is modelled as transversely isotropic rings in series. One end of the DNA is fixed, while a concentrated load, $F$, is applied to the other end.

There are a total of eight basic DNA sequence invariants based on structural symmetry [21], including AA/TT, CC/GG, AT/TA, CG/GC, ACCA/GTTG, AGGA/CTTC, ATGACA/ATCTGT, and ACCGGA/TGCGCT. In our coarse-grained model, we represent each invariant within short DNAs as a transversely isotropic ring with different material properties. For simplicity, we will initially focus on short DNAs composed solely of AT/TA and CG/GC base pairs. This is based on the recognition that DNAs constructed with varying AT/TA or CG/GC content have been found to exhibit significant differences in bending rigidity [23].

As shown in Figure 1a, we model the AT/TA or CG/GC sequence as a transversely isotropic elastic ring with the same geometry. The coordinates are defined in Fig. 1b, where axes "1" and "2" lie within the isotropic plane of the rings, and axis "3" represents the axial direction of the rings. In general, a transversely isotropic elastic material requires five independent material parameters, including elastic moduli $E_1$ and $E_3$, Poisson ratios $v_{12}$ and $v_{23}$, and shear modulus $G_{13}$. However, for the sake of simplicity, we assume $v_{12} = v_{23} = 0$ in our model, resulting in only three independent material parameters for each ring. It is important to note that the shear modulus $G_{13}$ of the rings may vary with salt concentration, and it can be relatively low under high salt concentration [24].

The persistence length of a short DNA, denoted as $l_p$, is related to its apparent bending modulus, denoted as $EI$, through $l_p = EI/kT$, with $k$ being the Boltzmann constant and $T$ the absolute temperature. To determine $l_p$ of a short DNA, we calculate its apparent bending modulus by treating the short DNA as an elastic beam. In our approach, a short DNA is fixed on one end and subjected to a concentrated force, denoted as $F$, on the other end, as illustrated in Fig. 1b. The contour length of the DNA is denoted as $L_c$. Utilizing the elastic beam theory, we can express the deflection of the beam at its free end, denoted as $w_F$, as $w_F = FL_c^3/3EI$. The deflection of $w_F$ can be obtained through simulation using software such as Abaqus. Subsequently, the persistence length of $l_p$ is determined by the equation, $l_p = FL_c^3/(3kTw_F)$.

Once the persistence length, $l_p$, is determined, we can further calculate the apparent $j$ factor, which has been measured in previous experiments [11, 19] and reflects the flexibility of short DNAs [11]. The apparent $j$ factor is calculated using a saddle point estimation of the partition function [25,26] by $j(r,L_c)=\int_0^r S(r',L_c)(dr')/((4\pi r^3)/3)$, where $r$ is the distance from end to end of a loop, $S$ is the radial probability density functions, given by $S(r',L_c) = 4\pi r'^2 Q(r',L_c)$, where $Q(r',L_c) = C_{SY}(L_c + r') \exp[-\Delta E(r',L_c)/(kT)]$, with $C_{SY}(L_c + r') = 1.66 \times 112.04 \times l_p^2/(L_c+ 2r')^5 \exp(0.246 \times (L_c + 2r')/l_p)$, and $\Delta E(r',L_c)/(kT) = 4 \times l_p/L_c \times \widehat{K}^2(x) \times (2x^2 - 1 + r'/L_c)$, where $x$ must satisfy $(1 + r'/L_c)\widehat{K}(x) = 2\widehat{E}(x)$ with $\widehat{K}(x)$ being the first complete elliptic integral and $\widehat{E}(x)$ being the second complete elliptic integral.

Default values for parameters used in our analysis are provided in the following. $D = 2$ nm [27], $d = 1.6$ nm, and $h = 2b$, with $b = 0.34$ nm, being the size of a base pair

[34]. $E_1 = 0.005$ GPa and $G_{13} = 0.001$ GPa. $E_3$ for AT/TA is 0.45 GPa and that for CG/GC is 0.9 GPa. Note that rings for AT/TA and CG/GC are exactly the same except that $E_3$ for AT/TA is lower than that for CG/GC [21].

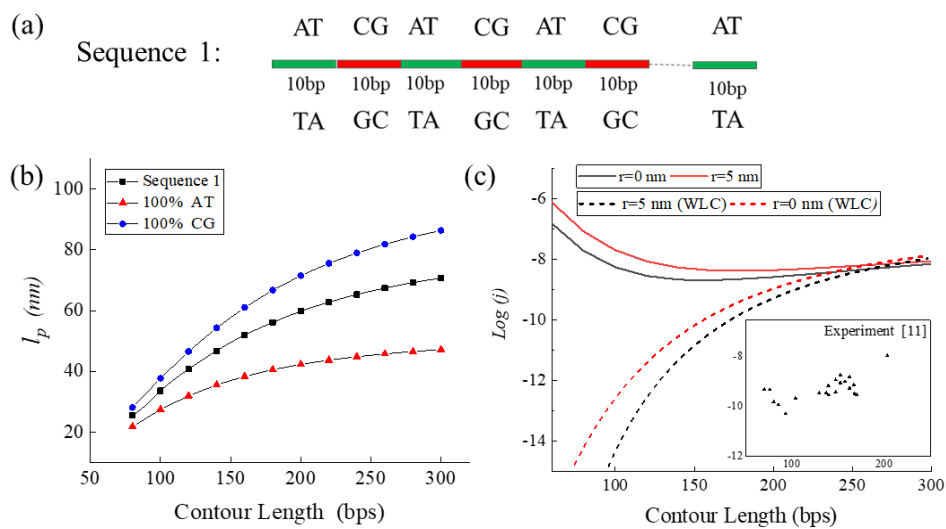

**Fig. 2** (a) In Sequence 1, AT/TA and CG/GC periodically appear every ten base pairs; (b) Variation of the persistence length of short DNAs with different sequences against the contour length; (c) Apparent $j$ factor for the looping of short DNAs: Solid lines are our model predictions for Sequence 1 and dashed lines are predictions based on the WLC. Inset shows the experimental data [11].

With the method described above, we investigate short DNAs with Sequence 1, where AT/TA and CG/GC sequences appear periodically every ten base pairs, as illustrated in Fig. 2a. Our model predictions, shown in Fig. 2b, indicate that the persistence length of short DNAs consistently increases with the contour length until it reaches saturation at a large length. The maximum value of the saturated persistence length is approximately 75 nm. From Fig. 2b, it can be inferred that for short DNAs with a contour length of 100 base pairs, the persistence length is less than half of 75 nm, indicating that very short DNAs can be highly flexible. Furthermore, we calculate the $j$ factors of short DNAs with Sequence 1, as displayed in Fig. 2c. It can be observed that the $j$ factor exhibits weak dependence on the contour length of short DNAs. As the contour length decreases, the corresponding *j* factor initially decreases and then increases, similar to the trends observed in experiments [11], where the looping rate of short DNAs was measured using a fluorescence-based and protein-free assay, with a weak length dependence. This is in contrast to the strong dependence of the *j* factor on the contour length of short DNAs predicted by the classical worm-like chain (WLC) theory, as also shown in Fig. 2c for comparison.

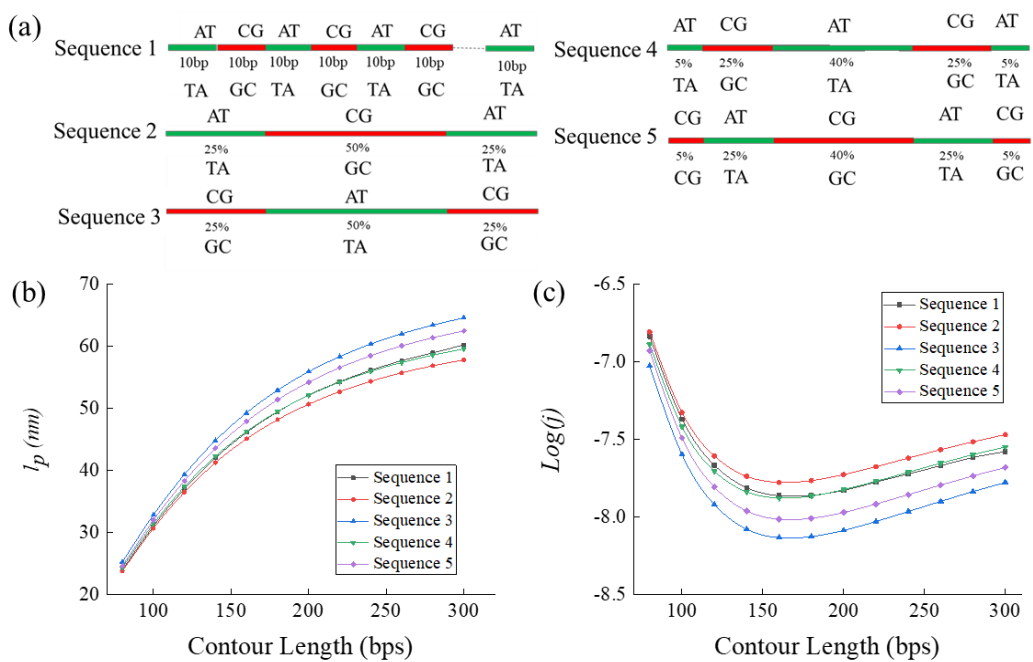

**Fig. 3** (a) Illustration of five short DNA sequences, each composed of a combination of 50% AT/TA and 50% CG/GC base pairs; (b) Variation of the persistence length of five short DNA sequences against the contour length; (c) Plotting the apparent *j* factors of five short DNA sequences against the contour length with *r*=5 nm.

To investigate the influence of DNA sequences on the persistence length of short DNAs in our model, we calculate the persistence length and apparent *j* factors for five different DNA sequences: Sequence 1, Sequence 2, Sequence 3, Sequence 4, and Sequence 5, as illustrated in Fig. 3a. All of these DNAs have the same AT/TA or CG/GC content, comprising 50% of the sequence. As evident from Figs. 3b,c, our analysis reveals that the persistence length of short DNAs is indeed affected by the DNA sequence. The maximum values of persistence length and *j* factor vary among the different DNA sequences, consistent with experimental findings [19]. In these experiments, different looping times were reported for different DNA sequences using a high-throughput assay, indicating the existence of a "mechanical code" with broad functional implications [19].

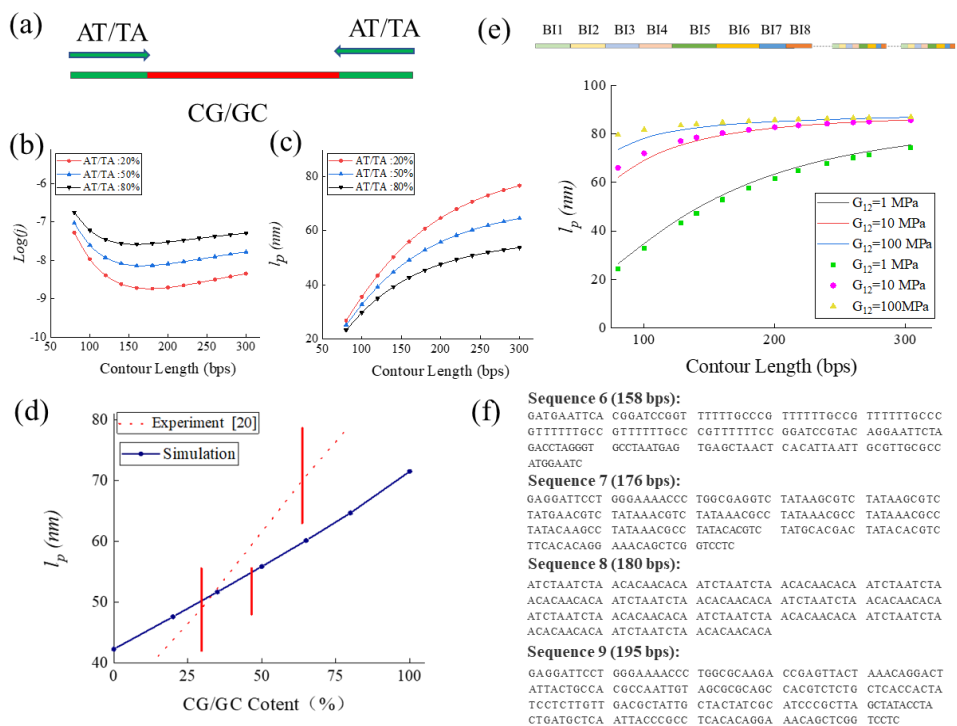

**Fig. 4** (a) Visual representation of DNA sequences featuring symmetrical distribution of AT/TA base pairs on both ends of short DNAs; (b) Plot depicting the variation of corresponding persistence length against the contour length; (c) Plot illustrating the variation of the corresponding apparent *j* factor against the contour length with $r = 5$nm; (d) The blue line represents the variation of short DNA persistence length against CG/GC content with $L_c$=68 nm. The red bars denote experimental results, with the dashed line indicating the trend [20]; (e) Variation of short DNA persistence length with eight basic invariant sequences against the contour length. Dots represent predictions utilizing the Timoshenko theory, while lines represent those based on finite element calculations. Inset: Illustration of a sequence containing eight basic invariants; (f) Four specific DNA sequences (Sequence 6, Sequence 7, Sequence 8, and Sequence 9) containing eight basic variants.

In order to study the effect of AT/TA content on the flexibility of DNA, we calculate the persistence length of short DNAs with a specific sequence, as illustrated in Fig. 4a. In this sequence, AT/TA sequences are symmetrically distributed on both ends of the short DNAs. We analyze the AT/TA content at different percentages, including 0%, 20%, 50%, 80%, and 100%. The results are shown in Figs. 4b,c. From Fig. 4b, it is evident that the persistence length of short DNAs increases with the contour length and decreases with increasing AT/TA content. The impact of AT/TA content on the persistence length of short DNAs with a contour length of 68 nm is also depicted in Fig. 4d, where the persistence length increases as the CG/GC content increases. Furthermore, we find that the values of the apparent $j$ factor are higher as the CG/GC content decreases in our analysis. This finding aligns with experimental results, where the persistence length of short DNAs with 30%, 47%, and 64% CG/GC content increased with CG/GC content [20]. It is in line with previous suggestions that DNA sequences rich in AT/TA content exhibit softer properties compared to average sequences in DNA looping [28].

By considering DNA sequences composed solely of AT/TA and CG/GC, our model predictions align significantly with experimental findings [11, 19, 29-36]. In natural short DNAs, there can be up to eight basic variants [21], namely AA/TT, CC/GG, AT/TA, CG/GC, ACCA/GTTG, AGGA/CTTC, ATGACA/ATCTGT, and ACCGGA/TGCGCT, denoted as BI1, BI2, BI3, BI4, BI5, BI6, BI7, and BI8, respectively. Hence, we employ eight distinct rings with different material parameters to model various DNA sequences found in nature. The precise selection of these material parameters requires further examination, potentially including considerations such as ion concentration [32] or temperature. As a tentative assignment, we assign different $E_3$ values to BI1-BI8, which are 0.3848 GPa, 0.9522 GPa, 0.45 GPa, 0.90 GPa, 0.9717 GPa, 1.0957 GPa, 0.7109 GPa, and 1.2717 GPa, respectively, based on their respective reported persistence lengths in the literature [21], which are 41.7 nm, 50.4 nm, 42.7 nm, 49.6 nm, 50.7 nm, 52.6 nm, 46.7 nm, and 55.3 nm, respectively. The outer diameter ($D$) and inner diameter ($d$) of these rings are equal, while the height ($h$) is set to the number of base pairs within each respective basic variant multiplied by $b$.

For demonstration, we present the model predictions of the persistence length for short DNAs with eight basic variants in Fig. 4e. As expected, the persistence length increases with the contour length, consistent with our previous findings. Moreover, we further investigate four specific DNA sequences containing the eight basic variants, as listed in Fig. 4f and referred to as Sequence 6, Sequence 7, Sequence 8, and Sequence 9. According to our predictions, the respective persistence lengths for these sequences are 57.0 nm, 53.3 nm, 60.5 nm, and 62.7 nm, which reasonably agree with the corresponding experimental data [35] of 52.5 nm, 51.5 nm, 58.2 nm, and 56.3 nm.

Interestingly, we have discovered that the deflection of our coarse-grained model for short DNAs can also be estimated using the classical Timoshenko beam theory, which incorporates shear effects on the cross-section of a beam. In accordance with this theory, the governing equations for the beam are given by $d^2w/dx^2 + d\Phi/dx = 0$ and $d^3\Phi/dx^3 = 0$, where $\Phi$ represents the rotation angle of the cross-section and $w$ represents the deflection of the beam. Using one ring to represent each basic variant within the short DNAs, we can express the deflection of the beam, $w_i$, and the rotation angle of the cross-section, $\Phi_i$, within the i$^{th}$ ring, as $w_i = a_i x^3 + b_i x^2 + c_i x + d_i$ and $\Phi_i = -3a_i x^2 - 2b_i x + e_i$. Here, $a_i$, $b_i$, $c_i$, $d_i$, and $e_i$ are constants to be determined using boundary conditions at the two ends of the cantilevered short DNAs, as shown in Fig. 1b, as well as continuity conditions between neighboring rings. By applying this alternative approach, we can obtain the deflection of the short DNAs at the free end. From this, we can calculate the persistence length using the same procedure described earlier.

As depicted in Fig. 4e, the predictions obtained with the classical Timoshenko beam theory demonstrate good agreement with those obtained through finite element calculations. Notably, only two elastic parameters, $E_3$ and $G_{12}$, are required in the Timoshenko beam theory, suggesting that $E_1$ may have a minor influence on our coarse-grained model. Furthermore, it is important to highlight that the dependence of the persistence length on the contour length of short DNAs is only significant when the shear modulus $G_{12}$ is much smaller than $E_3$. As illustrated in Fig. 4e, as $G_{12}$ increases, this length dependence becomes weaker.

In conclusion, we have introduced a transversely isotropic ring model to analyze the persistence length and apparent $j$ factors of short DNAs with varying contour length and sequences. Our findings demonstrate that the persistence length of short DNAs increases with contour length until it reaches saturation, and that the DNA sequence significantly influences the flexibility of short DNAs, aligning well with experimental observations. We highlight the crucial role of structural heterogeneity and anisotropy in determining the flexibility of short DNAs, which can be effectively captured by the classical Timoshenko beam theory incorporating shear considerations.

**Acknowledgements**

This work was supported by Zhejiang Provincial Natural Science Foundation of China (Grant No.: LZ23A020004) and the National Natural Science Foundation of China (Grant No.: 11872334).

**References**

[1] A. Cournac and J. Plumbridge, *DNA Looping in Prokaryotes: Experimental and Theoretical Approaches*, J. Bacteriol. **195**, 1109 (2013).

[2] J. Shin and A. B. Kolomeisky, *Facilitation of DNA Loop Formation by Protein-DNA Non-Specific Interactions*, Soft Matter **15**, 5255 (2019).

[3] J. S. Godde, S. U. Kass, M. C. Hirst, and A. P. Wolffe, *Nucleosome Assembly on Methylated CGG Triplet Repeats in the Fragile X Mental Retardation Gene 1 Promoter*, J. Biol. Chem. **271**, 24325 (1996).

[4] J.-F. Allemand, S. Cocco, N. Douarche, and G. Lia, *Loops in DNA: An Overview of Experimental and Theoretical Approaches*, Eur. Phys. J. E Soft Matter Biol. Phys. **19**, 293 (2006).

[5] Y. C. Fung, *Biomechanics*, Appl. Mech. Rev. **38**, 1251 (1985).

[6] R. J. Lang, K. A. Tolman, E. B. Crampton, S. P. Magleby, and L. L. Howell, *A Review of Thickness-Accommodation Techniques in Origami-Inspired Engineering*, Appl. Mech. Rev. **70**, 010805 (2018).

[7] C. Bustamante, J. F. Marko, E. D. Siggia, and S. Smith, *Entropic Elasticity of λ-Phage DNA*, Sci. Am. Assoc. Adv. Sci. **265**, 1599 (1994).

[8] J. F. Marko and E. D. Siggia, *Stretching DNA*, Macromolecules **28**, 8759 (1995).

[9] Q. Du, C. Smith, N. Shiffeldrim, M. Vologodskaia, and A. Vologodskii, *Cyclization of Short DNA Fragments and Bending Fluctuations of the Double Helix*, Proc. Natl. Acad. Sci. **102**, 5397 (2005).

[10] T. E. Cloutier, J. Widom, and R. D. Kornberg, *DNA Twisting Flexibility and the Formation of Sharply Looped Protein-DNA Complexes*, Proc. Natl. Acad. Sci. - PNAS **102**, 3645 (2005).

[11] R. Vafabakhsh and T. Ha, *Extreme Bendability of DNA Less than 100 Base Pairs Long Revealed by Single-Molecule Cyclization*, Sci. Am. Assoc. Adv. Sci. **337**, 1097 (2012).

[12] P. A. Wiggins, R. Phillips, and P. C. Nelson, *Exact Theory of Kinkable Elastic Polymers*, Phys. Rev. E Stat. Nonlin. Soft Matter Phys. **71**, 021909 (2005).

[13] T. E. Cloutier and J. Widom, *Spontaneous Sharp Bending of Double-Stranded DNA*, Mol. Cell **14**, 355 (2004).

[14] J. Yan and J. F. Marko, *Localized Single-Stranded Bubble Mechanism for Cyclization of Short Double Helix DNA*, Phys. Rev. Lett. **93**, 1 (2004).

[15] P. C. Nelson, P. A. Wiggins, T. van der Heijden, F. Moreno-Herrero, A. Spakowitz, R. Phillips, J. Widom, and C. Dekker, *High Flexibility of DNA on Short Length Scales Probed by Atomic Force Microscopy*, Nat. Nanotechnol. **1**, 137 (2006).

[16] A. Vologodskii and M. D. Frank-Kamenetskii, *Strong Bending of the DNA Double Helix*, Nucleic Acids Res. **41**, 6785 (2013).

[17] B. Chen and C. Dong, *Modeling Deoxyribose Nucleic Acid as an Elastic Rod Inlaid With Fibrils*, J. Appl. Mech. **81**, np (2014).

[18] D. Bhattacharyya, S. Kundu, A. R. Thakur, and R. Majumdar, *Sequence Directed Flexibility of DNA and the Role of Cross-Strand Hydrogen Bonds*, J. Biomol. Struct. Dyn. **17**, 289 (1999).

[19] A. Basu et al., *Measuring DNA Mechanics on the Genome Scale*, Nat. Lond. **589**, 462 (2021).

[20] M. J. Shon, S.-H. Rah, and T.-Y. Yoon, *Submicrometer Elasticity of Double-Stranded DNA Revealed by Precision Force-Extension Measurements with Magnetic Tweezers*, Sci. Adv. **5**, eaav1697 (2019).

[21] S. Geggier, A. Vologodskii, and M. Levitt, *Sequence Dependence of DNA Bending Rigidity*, Proc. Natl. Acad. Sci. **107**, 15421 (2010).

[22] D. Chakraborty, N. Hori, and D. Thirumalai, *Sequence-Dependent Three Interaction Site Model for Single- and Double-Stranded DNA*, J. Chem. Theory Comput. **14**, 3763 (2018).

[23] A. Kandinov, K. Raghunathan, and J.-C. Meiners, *Using DNA Looping to Measure Sequence Dependent DNA Elasticity*, in *Optical Trapping and Optical Micromanipulation IX*, Vol. 8458 (SPIE, 2012), pp. 154–162.

[24] F. Kriegel, N. Ermann, R. J. G. Forbes, D. Dulin, N. H. Dekker, and J. Lipfert, *Probing the Salt Dependence of the Torsional Stiffness of DNA by Multiplexed Magnetic Torque Tweezers*, Nucleic Acids Res. **45**, 5920 (2017).

[25] J. Shimada and H. Yamakawa, *Ring-Closure Probabilities for Twisted Wormlike Chains. Application to DNA*, Macromolecules **17**, 689 (1984).

[26] N. Douarche and S. Cocco, *Protein-Mediated DNA Loops: Effects of Protein Bridge Size and Kinks*, Phys. Rev. E Stat. Nonlin. Soft Matter Phys. **72**, 061902 (2005).

[27] R. R. Sinden, editor , *Subject Index*, in *DNA Structure and Function* (Academic Press, San Diego, 1994), pp. 393–398.

[28] S. Johnson, Y.-J. Chen, and R. Phillips, *Poly(DA:DT)-Rich DNAs Are Highly Flexible in the Context of DNA Looping*, PloS One **8**, e75799 (2013).

[29] D. Haldar and C. Schmuck, *Metal-Free Double Helices from Abiotic Backbones*, Chem. Soc. Rev. **38**, 363 (2009).

[30] J.-Y. Kim, J.-H. Jeon, and W. Sung, *A Breathing Wormlike Chain Model on DNA Denaturation and Bubble: Effects of Stacking Interactions*, J. Chem. Phys. **128**, 055101 (2008).

[31] T. M. Okonogi, S. C. Alley, A. W. Reese, P. B. Hopkins, and B. H. Robinson, *Sequence-Dependent Dynamics of Duplex DNA: The Applicability of a Dinucleotide Model*, Biophys. J. **83**, 3446 (2002).

[32] C. G. Baumann, S. B. Smith, V. A. Bloomfield, and C. Bustamante, *Ionic Effects on the Elasticity of Single DNA Molecules*, Proc. Natl. Acad. Sci. **94**, 6185 (1997).

[33] M. D. Frank-Kamenetskii, A. V. Lukashin, V. V. Anshelevich, and A. V. Vologodskii, *Torsional and Bending Rigidity of the Double Helix from Data on Small DNA Rings*, J. Biomol. Struct. Dyn. **2**, 1005 (1985).

[34] R. E. Franklin and R. G. Gosling, *Molecular Configuration in Sodium Thymonucleate*, Nat. Lond. **421**, 400 (2003).

[35] J. S. Mitchell, J. Glowacki, A. E. Grandchamp, R. S. Manning, and J. H. Maddocks, *Sequence-Dependent Persistence Lengths of DNA*, J. Chem. Theory Comput. **13**, 1539 (2017).

[36] S. Guilbaud, L. Salomé, N. Destainville, M. Manghi, and C. Tardin, Dependence of DNA Persistence Length on Ionic Strength and Ion Type, Phys. Rev. Lett. 122, 028102 (2019).